# Spin pumping using an $Ni_{80}Fe_{20}$ thin film annealed in a magnetic field

Hideki Shimogiku [a], Naoyuki Hanayama [b], Yoshio Teki [c], Hiroaki Tsujimoto [a], Eiji Shikoh [a,] *

[a] Graduate School of Engineering, Osaka City University, 3-3-138 Sugimoto, Sumiyoshi-ku, Osaka 558-8585, Japan

[b] Faculty of Engineering, Osaka City University, 3-3-138 Sugimoto, Sumiyoshi-ku, Osaka 558-8585, Japan

[c] Graduate School of Science, Osaka City University, 3-3-138 Sugimoto, Sumiyoshi-ku, Osaka 558-8585, Japan

*Corresponding author.

Tel.: +81 6 6605 2690; fax.: +81 6 6605 2690.

E-mail address: shikoh@elec.eng.osaka-cu.ac.jp (E. Shikoh).

Postal address: Graduate School of Engineering, Osaka City University, 3-3-138 Sugimoto, Sumiyoshi-ku, Osaka 558-8585, Japan

Abstract:

Spin pumping controlled with the ferromagnetic resonance of an $Ni_{80}Fe_{20}$ thin film annealed in a magnetic field was performed in order to investigate the simple and efficient generation method of the pure spin current. At the spin-pumping using the $Ni_{80}Fe_{20}$ on an annealed Pd/$Ni_{80}Fe_{20}$ stacked structure, the electromotive force due to the inverse spin-Hall effect (ISHE) in the Pd was found to be 30% stronger than that without annealing. When the angle between the directions of localized magnetic moments in the $Ni_{80}Fe_{20}$ film and the external magnetic field in the spin-pumping is zero, the spin injection efficiency into the Pd layer, i.e., the spin current density generated in the Pd layer can be the maximum. The annealing in a magnetic field is a convenient technique for increasing the spin current density generated by the spin pumping.

1. Introduction

The efficient generation and steady manipulation of the pure spin current will enable a new type of energy-saving electronic devices because the spin current carries only the spin degree of freedom, and not the charge degree of freedom, where the net charge flow is almost zero, and very little electric power is consumed [1]. To date, the pure spin current has generally been generated using electrical, dynamical, and thermal methods. Spin pumping, which is a dynamical method of generating the pure spin current, is a simple and powerful tool for research on pure spin currents [2-10], and the ferromagnetic metal $Ni_{80}Fe_{20}$, in which the magneto-crystalline anisotropy is negligible, has been widely used as the spin source on the spin pumping controlled with the ferromagnetic resonance (FMR) [3-6]. To increase the generation efficiency of the spin current using the spin pumping, other ferromagnetic materials such as $Fe_3Si$ [7], $Fe_3N$ [8], and $Co_2MnSi$ [9], which are well spin-polarized materials, have also been used because the spin current generated by the spin pumping of these materials has been expected to be well spin-polarized. To form thin films of such alloys, an epitaxial growth system based on molecular beam epitaxy, or reactive sputtering is normally used, however, which are relatively-expensive tools, and it is hard to form the films with the high quality. In this study, in order to investigate the creation of a pure spin current more simply and efficiently than by the previous materials and methods, the $Ni_{80}Fe_{20}$ is refocused because it's easily formed using conventional electron-beam (EB) deposition or sputtering methods, and the quality of the Ni$_{80}$Fe$_{20}$ films can be improved by annealing in a magnetic field, originating from the induced magnetic anisotropy [11]. Spin pumping using the Ni$_{80}$Fe$_{20}$ thin films annealed in a magnetic field was performed and the spin current with a higher density than that without annealing was successfully generated.

2. Material and methods

Figure 1 shows a schematic illustration of the stacked sample and the measurement method in this study. A pure spin current is generated in a palladium (Pd) layer by the spin-pumping of the Ni$_{80}$Fe$_{20}$ and is detected as the electromotive force in the Pd due to the inverse spin-Hall effect (ISHE) [4]. The Pd layer (10 nm thick) was formed on a thermally oxidized silicon substrate using an EB deposition system under a vacuum of $1.0 \times 10^{-6}$ Pa. Next, an Ni$_{80}$Fe$_{20}$ layer (25 nm thick) was formed using an EB deposition system with a shadow mask under a vacuum of $1.0 \times 10^{-6}$ Pa. Subsequently, without breaking the vacuum, aluminum (Al) was deposited (5 nm thick, not illustrated in Fig. 1) in order to prevent oxidation of the Ni$_{80}$Fe$_{20}$ layer. In this study, both as-deposited Ni$_{80}$Fe$_{20}$ films and Ni$_{80}$Fe$_{20}$ films annealed in a magnetic field after deposition were prepared for the comparison. Thus, after the all depositions, a certain amount of the samples were annealed under a vacuum of $1.0 \times 10^{-3}$ Pa, in a magnetic field of 30 mT, and the direction of the applied magnetic field during annealing was varied among the samples to confirm the annealing effect. The maximum annealing temperature used was set to be 400°C in order to avoid melting the Al or to avoid mixing of the Al and $Ni_{80}Fe_{20}$ at the interlayer. For control experiments, samples with a Cu film instead of a Pd film were investigated. Magnetic properties were measured using a vibrating sample magnetometer (TOEI, CV-300-5). The spin pumping behavior using the FMR and the ISHE were evaluated by the system composed of an electron spin resonance (ESR) system with a $TE_{011}$ cavity (JEOL, JES-TE300) and a nano-voltmeter (Keithley, 2182A) to detect the electromotive force in the samples. All measurements were performed at room temperature.

3. Results and discussion

Figure 2 shows magnetization curves for the samples. The black, red, and blue circles and lines represent the data for samples that were not annealed (hereafter, NA), annealed samples for which the static magnetic field directions for annealing and measurement were parallel (hereafter, AX), and annealed samples for which the static magnetic field directions for annealing and measurement were mutually orthogonal (hereafter, AY), respectively. Typical ferromagnetic properties were observed for the $Ni_{80}Fe_{20}$ films in all measurements. The squareness ratio of the magnetization curve for the AX sample is the best of all measurements.

Compared AX with AY, the anisotropic magnetic field $H_k$ was estimated to be 0.7 mT. These are evidence that magnetic anisotropy was induced in the $Ni_{80}Fe_{20}$ films.

Figure 3(a) shows FMR spectra near the ferromagnetic resonance field $H_{FMR}$ for the $Ni_{80}Fe_{20}$ films at a microwave power of 80 mW and the frequency $f$ of 9.45 GHz. The black, red, and blue lines represent the spectra of NA, AX, and AY samples, respectively. The $H_{FMR}$ values are 103 mT for NA, 109 mT for AX, and 122 mT for AY, respectively. The linewidth $W$ of the spectrum (shown in the figure) for the AX samples (4.21 mT) and that of the AY samples (6.10 mT) were larger than that for the NA sample (3.55 mT), which indicates that the magnetic anisotropy of the $Ni_{80}Fe_{20}$ has been changed. The $W$ difference between the AX and AY might be due to the difference of the magnetic anisotropy induced by the annealing.

Figure 3(b) shows the output voltages from NA and AX samples under a microwave power of 80 mW, for angles between the direction of the external magnetic field and the sample film plane $\theta$ of 0° and 180°. The red (black) open circles show experimental data from AX (NA) samples. The red (black) lines are the fitting results for AX (NA) samples, which were obtained using the following equation [4-6]:

$$V(H) = V_{ISHE} \frac{\Gamma^2}{(H-H_{FMR})^2 + \Gamma^2} + V_{Asym} \frac{-2\Gamma(H-H_{FMR})}{(H-H_{FMR})^2 + \Gamma^2}, \qquad (1)$$

where $\Gamma$ denotes the damping constant (3.20 mT for NA, 3.50 mT for AX). The first and second terms in Eq. (1) correspond to contributions from the ISHE and the asymmetry term against $H$ (i.e., anomalous Hall effect [4]), respectively. $V_{ISHE}$ and $V_{Asym}$ correspond to the coefficients of the first and second terms in Eq. (1). In both samples containing a Pd film, the output voltage was observed near $H_{FMR}$, and the signs of the output signals for the $\theta$ of 0° and 180° were opposite. This sign symmetry is a typical behavior that the obtained output voltage originates from the ISHE [4]. Meanwhile, no output voltage was observed for samples with a Cu film instead of a Pd film. Therefore, the obtained electromotive force in samples with a Pd film originates only from the ISHE in the Pd, and other extrinsic effects such as the self-induced ISHE of the $Ni_{80}Fe_{20}$ layer [12], the anomalous Nernst effect of the $Ni_{80}Fe_{20}$ [13], and the galvanic effect [10] are excluded. Using Eq. (1), the electromotive force due to the ISHE for Pd was estimated to be 1.93 μV for NA and 2.59 μV for AX, at $\theta = 0°$. That is, after annealing in an external magnetic field, the electromotive force due to the ISHE in the Pd was 30% larger than that without annealing. Meanwhile, the $V_{ISHE}$ for AY (not plotted in Fig. 3(b) to avoid confusion) was 1.88 μV, which was comparable to that for NA.

Using the experimental data for $\theta = 0°$, the spin current density $j_s$ generated in the Pd was estimated as follows: Using the obtained $V_{ISHE}$ for the Pd, the spin-Hall angle $\theta_{SHE}$ for the Pd of 0.01 [5], the spin diffusion length in the Pd $\lambda_{Pd}$ of 9 nm [14], the length of the $Ni_{80}Fe_{20}$ $w$ of 1,000 μm, the thickness of the Pd layer $d_{Pd}$ of 10 nm, the conductivity of the Pd, $\sigma_{Pd}$, and the conductivity of the $Ni_{80}Fe_{20}$, $\sigma_F$, the $j_s$ can be approximated by [5]

$$j_s = \frac{d_{Pd}\sigma_{Pd} + d_F\sigma_F}{w\theta_{SHE}\lambda_{Pd}\tanh(d_{Pd}/2\lambda_{Pd})}(\frac{\hbar}{2e})V_{ISHE}. \qquad (2)$$

From the Eq. (2), the $j_s$ values for the NA, AX and AY samples are estimated to be $7.94\times10^{-10}$ $10.7\times10^{-10}$ J/m$^2$, and $7.73\times10^{-10}$ J/m$^2$, respectively. Thus, the AX condition is the best among the samples. It is also known that the $j_s$ can be estimated using the damping constant $\alpha$ for the Ni$_{80}$Fe$_{20}$, the saturation magnetization $M_s$ for the Ni$_{80}$Fe$_{20}$, and the real part of the spin mixing conductance $g_r^{\uparrow\downarrow}$ at the interface between the Ni$_{80}$Fe$_{20}$ and Pd layers [5, 6]. For example, the $\alpha$ are estimated to be 0.096 for NA, 0.114 for AX, and 0.165 for AY using the obtained $W$ data [5]. However, the magnetic anisotropy has been intentionally induced in this study. As the result, the $W$ has includes not only the spin-pumping effect, but also the magnetic anisotropy effect by the annealing. Therefore, the second calculation method for the $j_s$ causes wrong values in this study.

We assume a model shown in figure 4, which explains the reason why only the AX sample has shown the maximum $V_{ISHE}$ and the values for AY and NA are comparable. Only the AX sample has the configuration without energy loss just before spin injection because the angle between the directions of localized magnetic moments in the Ni$_{80}$Fe$_{20}$ film and the external magnetic field in the spin-pumping is nearly zero, due to the induced magnetic anisotropy by the annealing. Meanwhile, for other two configurations (NA and AY), the angles between the directions of magnetic moments in the Ni$_{80}$Fe$_{20}$ film and the external magnetic field in the spin-pumping are finite values. That is, to align the orientation of the spin polarized vector of the spin current, the localized magnetic moments must rotate to the same direction as the direction of the static magnetic field, which cause an energy-loss before the spin injection into the Pd layer. As the result, the spin current density generated in the Pd layer will become low. It is also possible that the quality of the $Ni_{80}Fe_{20}$/Pd interface changes during the annealing and as a result, the exchange coupling at the interface may be improved by the annealing. However, if the interface quality were dominant, the AY sample would have shown an increase of the electromotive force. At the current stage, although there is no clear reason why the AX sample shows larger $V_{ISHE}$ than those for the NA and AY samples, the most important thing is that the $j_s$ for the AX sample, which the static magnetic field directions for annealing and measurement were parallel was larger than other two samples, and the electromotive force was increased by annealing the $Ni_{80}Fe_{20}$ layer. This suggests that the annealing in a magnetic field is a convenient technique for increasing the spin current density by the spin pumping, and useful for the development of novel spintronic devices.

4. Conclusions

Spin pumping controlled with the ferromagnetic resonance of the $Ni_{80}Fe_{20}$ in Pd/$Ni_{80}Fe_{20}$ samples annealed in a magnetic field was successfully performed. When the magnetic field direction during the measurements was the same as that during annealing, the electromotive force due to the ISHE in the Pd was found to be 30% larger than that without annealing. When the angle between the directions of magnetic moments in the $Ni_{80}Fe_{20}$ film and the external magnetic field in the spin-pumping is zero, the spin injection efficiency into the Pd layer, i.e., the spin current density generated in the Pd layer can be the maximum. The annealing effect in a magnetic field was also confirmed by the magnetization curves of the samples. Annealing in a magnetic field is a convenient technique for increasing the spin current density generated by the spin pumping.

Acknowledgements

This research was supported in part by a Grant-in-Aid for Scientific Research from the MEXT, Japan, and the Murata Science Foundation, Japan.

Figure captions

Fig. 1. (Color online) Schematic illustration of the stacked sample and the measurement method.

Fig. 2. (Color online) Magnetization curves for samples at room temperature. The black, red, and blue circles and lines represent the data from samples without annealing (NA), annealed samples for which the static magnetic field directions during annealing and the measurement were parallel (AX), and annealed samples for which the static magnetic field directions during annealing and measurement were mutually orthogonal (AY), respectively.

Fig. 3. (Color online) (a) FMR spectra at a microwave power of 80 mW and the frequency of 9.45 GHz. The black, red, and blue lines represent the spectrum for an NA sample, an AX sample, and an AY sample, respectively. (b) Output voltages from NA and AX samples under a microwave power of 80 mW, and at angles $\theta$ of 0° and 180°. The red (black) open circles represent experimental data from AX (NA) samples. The red (black) lines show the theoretical fitting of the AX (NA) sample data using Eq. (1).

Fig. 4. (Color online) Schematic illustration of the model to explain the reason why only the AX

sample has shown the maximum $V_{\text{ISHE}}$ and the values for AY and NA are comparable. White thick arrows correspond to localized magnetic moments in the $Ni_{80}Fe_{20}$ films and the orientations after the annealing (except for NA). Dashed arrows indicates the static field direction at the ISHE measurements of the $\theta = 0°$. White thin arrows correspond the motion of the localized magnetic moments when the static magnetic field is applied.

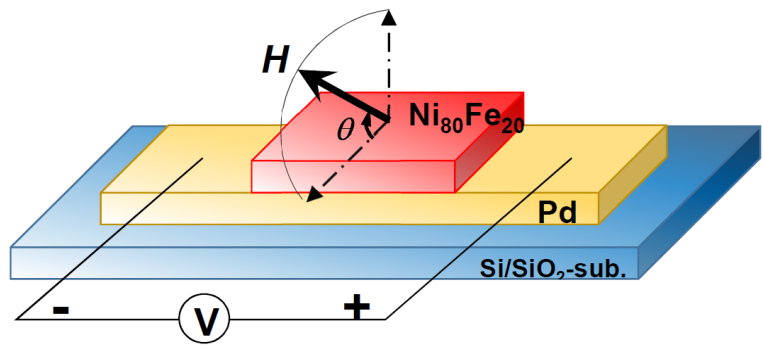

H. Shimogiku, *et al.*,   Fig. 1.

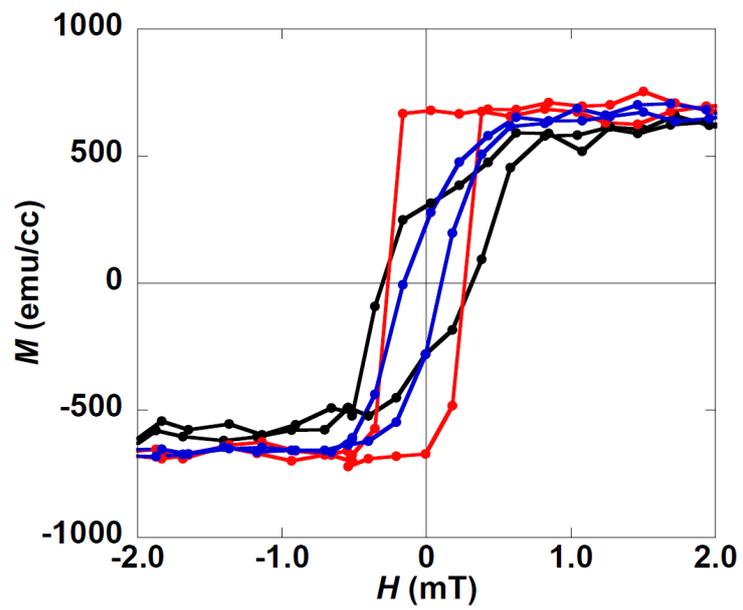

H. Shimogiku, *et al.*,    Fig. 2.

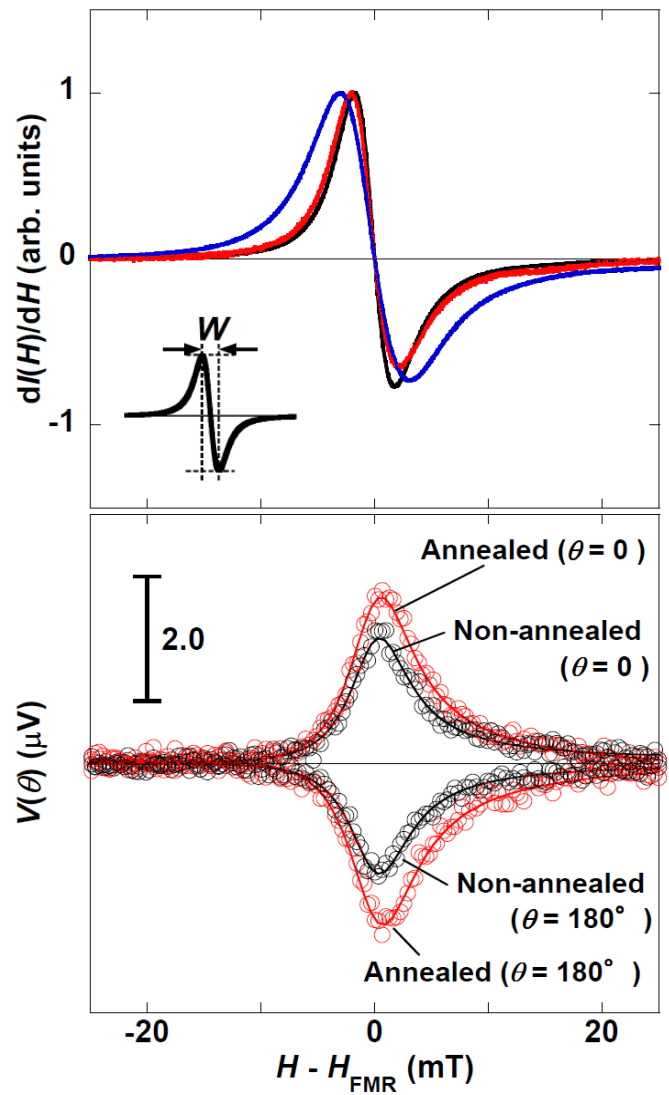

H. Shimogiku, *et al.*, Fig. 3.

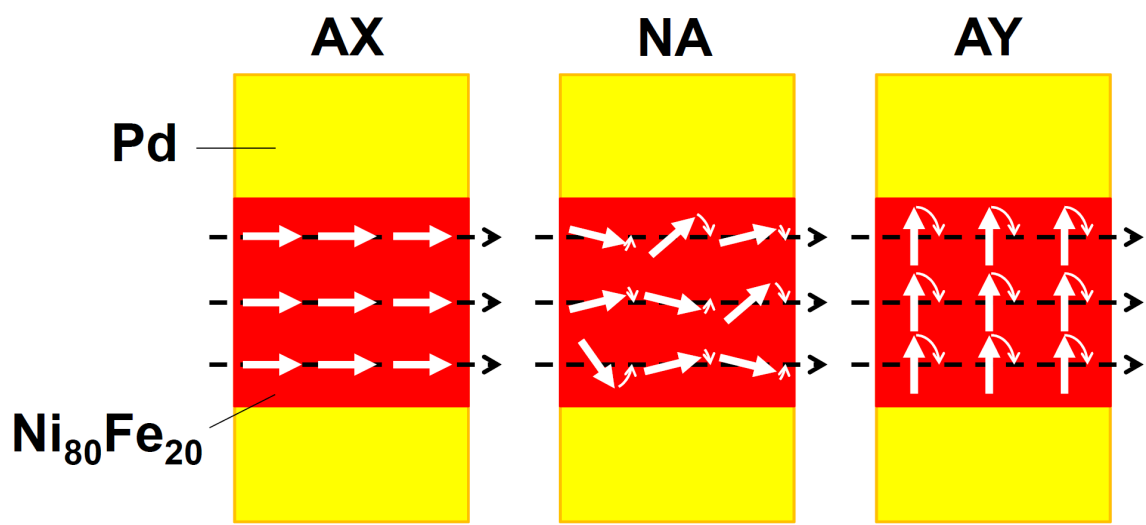

H. Shimogiku, *et al.*,   Fig. 4.